\documentclass[prl,aps,twocolumn, floatfix, superscriptaddress,showpacs]{revtex4}
\usepackage{amsmath,bm,graphicx}
\usepackage{amssymb}
\usepackage{amsfonts}
\usepackage{times}
\usepackage{graphicx,color}
\definecolor{r}{rgb}{1,0,0}
\definecolor{b}{rgb}{0,0,1}

\begin{document}

\title{Heat transport in the geostrophic regime of rotating Rayleigh-B{\'e}nard convection}

\author{Robert E. Ecke}
\affiliation{Center for Nonlinear Studies, Los Alamos National Laboratory, Los Alamos, NM 87545, USA}

\author{Joseph J. Niemela}
\affiliation{International Center for Theoretical Physics, Strada Costiera 11, 34014 Trieste, Italy}

\email[Corresponding Author: ]{ecke@lanl.gov}

\date{\today}

\begin{abstract}
We report experimental measurements of heat transport in rotating Rayleigh-B{\'e}nard convection in a cylindrical convection cell with aspect ratio $\Gamma = 1/2$. The fluid was helium gas with Prandtl number Pr = 0.7. The range of control parameters was Rayleigh number $4 \times 10^9 < {\rm Ra} < 4 \times 10^{11}$ and Ekman number $2 \times 10^{-7} < {\rm Ek} < 3 \times 10^{-5}$(corresponding to Taylor number $4 \times 10^9 < {\rm Ta} < 1 \times 10^{14}$  and convective Rossby number $0.07 < {\rm Ro} < 5$). We determine the crossover from weakly rotating turbulent convection to rotation dominated geostrophic convection through experimental measurements of the normalized heat transport Nu. The heat transport for the rotating state in the geostrophic regime, normalized by the zero-rotation heat transport, is consistent with scaling of $({\rm RaEk}^{-7/4})^\beta$ with $\beta \approx 1$. A phase diagram is presented that encapsulates measurements on the potential geostrophic turbulence regime of rotating thermal convection.
\end{abstract}
\maketitle

Thermal convection in the presence of rotation occurs in many geophysical contexts, including the Earth's mantle \cite{Glatzmaier1999}, oceans \cite{Marshall1999}, planetary atmospheres such as Jupiter \cite{HeimpelNature2005}, and solar interiors \cite{Spiegel1971}.  It also remains a fundamental problem in fluid dynamics, balancing rotation and buoyancy in a simple system that can be studied theoretically \cite{chandra61}, experimentally \cite{Rossby69,Zhong93,LE-PRL97,KSNHA09,ZSCVLA-PRL09,LE-PRE09} and numerically \cite{Julien-JFM96,SZCAL-PRL09} with high precision.  Thus, the problem of rotating thermal convection is of interest across a wide spectrum of scientific disciplines.

The parameters of rotating convection are ${\rm Ra} = {\rm g} \alpha \Delta {\rm T} d^3/\nu\kappa$ which measures the buoyant forcing of the flow, Ek = $\nu/(2 d^2 \Omega)$ which represents an inverse dimensionless rotation rate, and Pr = $\nu/\kappa$, where $g$ is acceleration of gravity, $\Delta$ T is the temperature difference between top and bottom plates separated by distance $d$, $\nu$ and $\kappa$ are the fluid kinematic viscosity and thermal diffusivity, respectively, and $\Omega = 2\pi f$ is the angular rotation about an axis parallel to gravity for rotation frequency $f$.  Rotation can also be represented by the Taylor number Ta = 1/Ek$^2$ or by the convective Rossby number Ro = Ek $\sqrt{{\rm Ra/Pr}}$ which reflects the ratio of rotational time to buoyancy time.  Here we use the representation of Ek or Ro such that high dimensionless rotation rates correspond to small values of the rotational control parameter in the spirit of the asymptotic equation approach of expanding in a small variable \cite{SpragueJFM2006}.  The measured response of the system in this space of buoyant and rotational forcing is the Nusselt number, Nu = $\dot{Q}/(\lambda \Delta {\rm T})$ where $\dot{Q}$ is the applied heater power through the fluid and $\lambda$ is the thermal conductance of the fluid. 

Much of the experimental work on rotating convection at high dimensionless rotation rates has focused on either the transition to convection where rotation-induced wall modes play an important role \cite{Pfotenhauer87,Zhong93} or the turbulent state far from onset where thermal boundary layers control heat transport \cite{Rossby69,LE-PRL97,ZSCVLA-PRL09,KSNHA09,LE-PRE09}. Recently, the numerical simulation 
\cite{GroomsPRL10,JulienPRL12} of the appropriate equations of motion \cite{SpragueJFM2006} in the asymptotic limit of high rotation rate has focused on the heat transport scaling above the convective onset but below the transition to boundary-layer controlled turbulence - a region that we will refer to as geostrophic turbulence.  Several predictions have been made for the scaling of heat transport in this regime with power laws in Rayleigh number, Ra, corresponding to Ra$^{3/2}$, based on numerical simulations \cite{JulienPRL12}, or Ra$^3$ based on dimensional arguments \cite{Boubnov90,KingJFM2012}. The data in this regime are scarce and the range of Prandtl number, Pr, Ekman number Ek (proportional to the inverse of the angular rotation rate $\Omega$), and Ra is very limited.  In particular, the crossovers from buoyancy dominated turbulent convection (where rotation has no measurable effect) to rotation-influenced turbulent convection (dominated by thermal 
boundary layer development) to geostrophic turbulence (Ek small) have not been well investigated.

\begin{figure}[h]
\begin{center}
\includegraphics[width = 3.2 in]{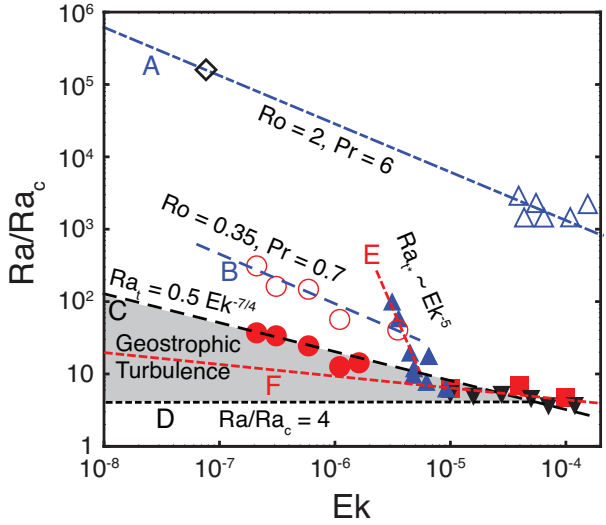}
\vspace{-0.5cm}
\end{center}
\caption{Phase diagram of rotating convection in parameters Ra/Ra$_c$ and Ek.  Rotation first affects turbulent convection below line A for Pr $= 6$ (Ro = 2) and below line B for Pr = 0.7 (Ro = 0.35). The crossover to geostrophic turbulence is roughly independent of Pr (or depends on it very weakly) and  occurs along line $C$ where ${\rm Ra_t} = 0.5 {\rm Ek}^{-7/4}$ and above line D which indicates Ra/Ra$_c =4$ corresponding roughly to a regime of weakly non-linear convection near onset.   Line E indicates a rapid change in the crossover for $Pr = 4.4$ corresponding to ${\rm Ra}_{t*} \sim {\rm Ek}^{-5}$. Line F is the upper limit of applicability of proposed ${\rm Ra}^3{\rm Ek}^4$ scaling corresponding to the relationship ${\rm Ra/Ra}_c \lesssim {\rm Ek}^{-1/6}$. Data are from \protect{\cite{KSNHA09}}(solid square - red), \protect{\cite{ZSCVLA-PRL09,ZhongJFM10}} (open and solid up triangles - blue), \protect{\cite{LE-PRE09}} (solid down triangles - black), \protect{\cite{NiemelaJFM2010}} (open diamond 
- black), and this work (open and solid circles - red). The shaded region corresponds to states expected to show geostrophic turbulence.
 }
 \label{fig:PhaseDiagram}
\end{figure}

The experimental apparatus used for these studies has been described in detail previously \cite{NiemelaNature2000,NiemelaJFM2010} so we include only essential details.  The convection cell had a cylindrical geometry with height $d$ = 100 cm and diameter 50 cm resulting in an aspect ratio $\Gamma = 1/2$.  The working fluid was helium gas near its critical point at around 5.2K, and the range of Ra and Ek was controlled by a combination of variations of $\Delta$ T in the range 0.04 - 0.30 K at a mean cell temperature between 4.61 and 4.75 K and densities $\rho$ of 0.00033, 0.00066, 0.0013, and 0.0018 g/cc.  The Prandtl number was constant at Pr = 0.7. For most of the runs, $f$ was fixed at the maximum for the apparatus corresponding to 0.167 Hz resulting in runs at constant Ek (and Ta).  In one run, Ra was fixed and $f$ varied between 0.0056 and 0.167 Hz.  For all of the data, Nu was measured without rotation as a reference and is denoted Nu$_0$.  The data are reported in standard ratios of Nu(Ra, $\Omega$)/Nu$_
0$(Ra, 0), that to first order compensate for small systematic uncertainties, and also facilitates comparison to other data sets by providing a self-reference for the system that takes into account the possibility of small differences in static and/or dynamic boundary conditions, etc.

We measure the convective heat transport through the Nusselt number Nu and explore the crossover to geostrophic rotating turbulence over a parameter range $2 \times 10^{-7} < {\rm Ek} < 3 \times 10^{-5}$ and $4 \times 10^9 < {\rm Ra} < 4 \times 10^{11}$, corresponding to a range of convective Rossby Number, $0.07 < Ro < 5$.  We find that the crossover to buoyancy dominated turbulence has a strong $Pr$ dependence whereas the crossover to geostrophic turbulence has, at most, a weak Pr dependence and, over a range of moderate Pr and more than 3 orders of magnitude in Ek, occurs for values of Ra consistent with ${\rm Ra}_t = 0.5 {\rm Ek}^{-7/4}$.  A summary of the resultant phase diagram based on a combination of our measurements with earlier measurements at larger Pr and Ek \cite{KSNHA09,ZSCVLA-PRL09,LE-PRE09} (mostly water at different mean temperatures) is shown in fig.\ \ref{fig:PhaseDiagram} where we normalize Ra by the rotation-dependent Ra$_c = 8.4 {\rm Ek}^{-4/3}$, 
representing linear stability values: in general, one needs Ra/Ra$_c \gg 1$ to achieve a strongly nonlinear turbulent state.  The strong Pr dependence of the crossover from buoyancy-dominated to rotation-influenced thermal boundary layer turbulence is easily demonstrated by comparing lines A and B.  Line C shows the extrapolation of our Ra$_t \sim {\rm Ek}^{-7/4}$ curve to include the higher Pr data.  The high Pr data are consistent with the Ek$^{-7/4}$ relationship for $8 \times 10^{-6} < {\rm Ek} <  \times 10^{-4}$.  On the other hand, the high Pr data for smaller Ek \cite{ZhongJFM10} show an abrupt increase along line E with Ra$_{t*} \sim {\rm Ek}^{-5}$, an apparently unnoticed feature of the high Pr data. We also show line F, the upper limit of self-consistency for arguments \cite{Boubnov86,KingJFM2012} leading to Nu $\sim$ Ra$^3$Ek$^4$, which corresponds to ${\rm Ra/Ra}_c \lesssim {\rm Ek}^{-1/6}$ which we derive below.  Finally, we find that the scaling of 
Nu with Ra 
in the geostrophic regime is consistent with a power law of order 1;  no evidence for power-law scaling of Nu $\sim {\rm Ra}^3$ \cite{KingJFM2012} is found. 

\begin{figure}[h]
\begin{center}
\includegraphics[width = 3.2 in]{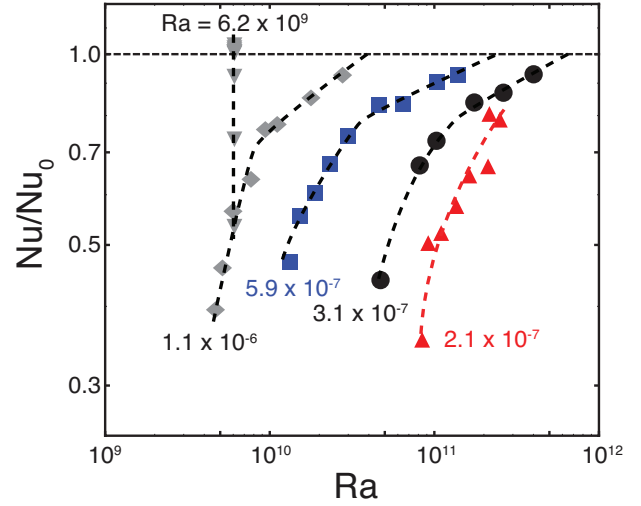}
\vspace{-0.5cm}
\end{center}
\caption{(Color Online) Nu/Nu$_0$ vs Ra for constant Ek: $1.1 \times 10^{-6}$ (solid diamond - gray),  $5.9 \times 10^{-7}$ (solid square - blue),  $3.1 \times 10^{-7}$ (solid circle - black),  $2.1 \times 10^{-7}$ (solid up triangle - red), and for constant Ra = $6.2 \times 10^9$ (solid down triangle - gray). Dashed lines are guides to the eye.
 }
 \label{fig:NuRa}
\end{figure}

As in earlier experiments \cite{NiemelaJFM2010} in this apparatus at much higher Ra and Pr = 6, Nu/Nu$_0 \le 1$ for all parameters measured as shown in fig.\ \ref{fig:NuRa}.  The data are for four different runs corresponding to constant Ek between $10^{-6}$ and $10^{-7}$ and one run at constant Ra = $6.2 \times 10^9$. From these data, we determine the Ek dependent values of Ra where Nu/Nu$_0$ drops below 1.  These transition values Ra$_T$ are plotted in fig.\ \ref{fig:PhaseDiagram} (line B) with an Ek dependence consistent with a constant ${\rm Ro} \approx 0.35$.  The data also suggest a second change in slope of the curves for smaller Ra but this behavior shows up more clearly if we scale the data so that they collapse onto a single curve.

\begin{figure}[h]
\begin{center}
\includegraphics[width = 3.2 in]{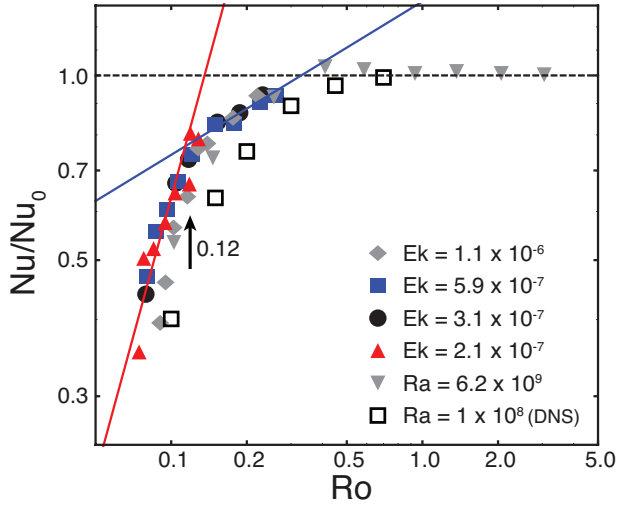}
\vspace{-0.5cm}
\end{center}
\caption{(Color Online) Log-Log plot of Nu/Nu$_0$ vs Ro for constant Ek: $1.1 \times 10^{-6}$ (solid diamond - gray),  $5.9 \times 10^{-7}$ (solid square - blue),  $3.1 \times 10^{-7}$ (solid circle - black),  $2.1 \times 10^{-7}$ (solid up triangle - red), and for constant Ra: $6.2 \times 10^9$ (solid down triangle - gray), DNS \protect{\cite{ZSCVLA-PRL09}} - $1 \times 10^8$ (open square - black). The solid lines show approximate power law variations of the region $0.8 < {\rm Nu/Nu}_0 < 1$  with Ro$^{1/4}$ (top, blue) and for ${\rm Nu/Nu}_0 < 0.8$ with Ro$^{3/2}$ (bottom, red), respectively.  Vertical arrow indicates approximate transition value Ro$=0.12$.
 }
 \label{fig:NuRo}
\end{figure}

One possibility for collapsing the data is to plot them in terms of Ro (proportional to Ra$^{1/2}$Ek  or equivalently Ra Ek$^2$) which we show in fig.\ \ref{fig:NuRo}. The collapse is reasonable although the DNS data \cite{ZSCVLA-PRL09} at much lower Ra are not captured well by this scaling.  The collapse does suggest two ranges of behavior consisting of an initial decrease in Nu/Nu$_0$ with decreasing Ro starting at Ro $\approx 0.35$ and a second more rapid decrease starting at Ro $\approx 0.1$. Nu/Nu$_0$ has dropped to about 0.8 at this second decrease.  The solid lines indicate power law curves corresponding to Ro$^{1/4}$ for the first decrease and Ro$^{3/2}$ for the faster decrease. No particularly strong conclusions can be drawn from these scalings but they are convenient for characterizing the form of the data. The lack of collapse of the DNS data at much lower Ra, however, anticipates that the scaling can be improved. 

\begin{figure}[h]
\begin{center}
\includegraphics[width = 3.2 in]{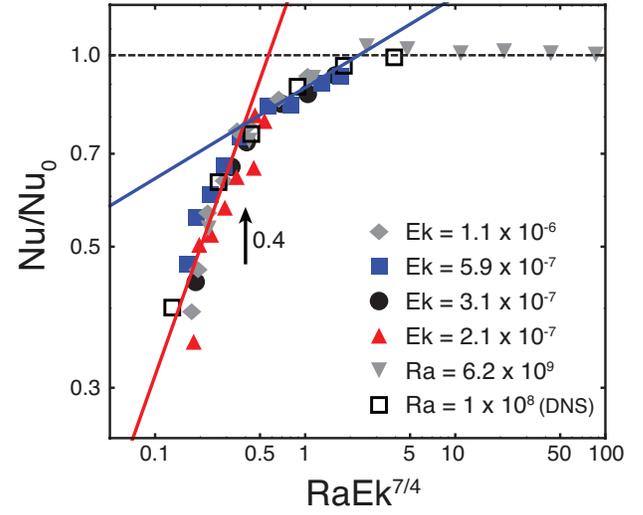}
\vspace{-0.5cm}
\end{center}
\caption{(Color Online) Log-Log plot of Nu/Nu$_0$ vs RaEk$^{7/4}$ for constant Ek: $1.1 \times 10^{-6}$ (solid diamond - gray),  $5.9 \times 10^{-7}$ (solid square - blue),  $3.1 \times 10^{-7}$ (solid circle - black),  $2.1 \times 10^{-7}$ (solid up triangle - red), and for constant Ra: $6.2 \times 10^9$ (solid down triangle - gray), DNS \protect{\cite{ZSCVLA-PRL09}} - $1 \times 10^8$ (open square - black). The solid lines show approximate power law variations of the region $0.8 < {\rm Nu/Nu}_0 < 1$  with 
$({\rm RaEk}^{7/4})^{1/7}$ (top, blue) and for ${\rm Nu/Nu}_0 < 0.8$ with $({\rm RaEk}^{7/4})^{2/3}$ (bottom, red), respectively. Vertical arrow indicates approximate transition value RaEk$^{7/4} = 0.4$, slightly less than the best value of 0.5 that fits all the data.
 }
 \label{fig:NuRaE74}
\end{figure}

Recently, measurements in water with Pr $\approx 6$ were conducted \cite{KSNHA09} in which the crossover between the boundary-layer dominated turbulent state and convection with Nu/Nu$_0 < 1$ was attributed to competing thermal and Ekman boundary layers.  The resulting crossover was found to have the form Ra$_t = 1.4 {\rm Ek}^{-7/4}$, suggesting the scaling variable Ra Ek$^{7/4}$.  We show the data normalized in this manner in fig.\ \ref{fig:NuRaE74}.  The collapse for our data is  better for this scaling with the DNS data now collapsed as well.  The difference in the two scalings is rather small, i.e., Ra Ek$^2$ in fig.\ \ref{fig:NuRo} compared to Ra Ek$^{7/4}$ in fig.\ \ref{fig:NuRaE74}.  It takes the large difference in Ra for the DNS to differentiate the two scalings.  A more recent analysis of the experimental water data \cite{KingJFM2012} suggested that a better fit was to Ra Ek$^{3/2}$ but that does not fit our data very well. Furthermore, the DNS results also are then 
significantly shifted to the left of the experimental points. Indeed, to within our experimental uncertainties, we find Ra$_t = 0.5 {\rm Ek}^{-7/4}$ as plotted in the phase diagram in fig.\ \ref{fig:PhaseDiagram}. Interestingly this relationship fits the crossover estimates for data with higher Pr in the range 3-6 \cite{KSNHA09,ZSCVLA-PRL09,LE-PRE09} and for Ek $ > 4 \times 10^{-6}$ but with a lower prefactor than the 1.4 suggested earlier \cite{KSNHA09}.

The power law straight lines in fig.\ \ref{fig:NuRaE74} are consistent with those in fig.\ \ref{fig:NuRo}, yielding relationships of Ra$^{1/7}$ and Ra$^{2/3}$ for the slopes.  Again these lines are drawn for the purposes of describing the data collapse and are over quite limited ranges of parameters.  There is, however, no evidence for Ra$^3$ scaling.  

Expanding on the limits for a geostrophic turbulence regime, numerical simulations \cite{JulienPRL12} suggest that one needs Ra/Ra$_c$ to be larger than about 4 to 6 to enter a regime of geostrophic turbulence. We denote this limit in fig.\ \ref{fig:PhaseDiagram} as line D.  One implication of this cutoff that can be drawn from the phase diagram is that experiments at higher Ek$ > 10^{-5}$ do not have a measurable range of geostrophic turbulence.  At the lower end for the present experiments, the range of Ra/Ra$_c$ available is ultimately limited by $\dot{Q}$ and $\Delta$T resolution.  In principle, one would like measurements of Nu over a range of Ra such that ${\rm Ra}_c << {\rm Ra} << {\rm Ra}_t$.  This limit suggests that one needs Ek$ < 10^{-7}$ to achieve a sufficient range to measure a decade of scaling of Nu with Ra in the geostrophic turbulence range and Ek$ < 10^{-9}$ to approach two decades.  This will be a stiff challenge for future experiments. In the present case, it is unclear whether our 
scaling of Nu/Nu$_0 \sim {\rm Ra}^\beta$ with $\beta \approx 1$ is obtained far enough below Ra$_t$ not to be influenced by that crossover; i.e., larger values of $\beta$ are not completely ruled out by these measurements. Nevertheless, the data presented here have the lowest Ek (highest dimensionless rotation rate) and largest range that resolves the geostrophic turbulence range of experiments performed until now.

Another surprise in our phase diagram is the very steep Ek dependence of the apparent crossover to boundary layer rotation-influenced turbulent convection.  These data were obtained in experiments \cite{ZhongJFM10} that were not designed to determine the low Ro crossover but instead concentrated on the crossover at higher Ra/Ra$_c$ to rotation-free buoyancy driven turbulence. Measurements on that system indicated that the aspect ratio $\Gamma$ plays a role in determining the upper boundary \cite{WeissJFM2011} and that there is a Pr dependence of the crossover \cite{ZhongJFM10} of approximately ${\rm Ro}_c \sim {\rm Pr}^{0.4}$.  On the first point, the data  at very high Ra in helium gas with $\Gamma = 0.5$ \cite{NiemelaJFM2010} suggest that the transition remains at ${\rm Ro}_c \approx 2$ independent of $\Gamma$ so there may be Ra dependence since an implication of the measured $\Gamma$ dependence \cite{WeissJFM2011}  would be ${\rm Ro}_c \approx 4$.  Second, if we take lines A and B as describing the 
data 
for Pr = 6 and Pr = 0.7, respectively, the implied Pr dependence would be ${\rm Ro}_c \sim {\rm Pr}^{0.8}$ rather than the previously indicated ${\rm Pr}^{0.4}$ dependence.  Finally, coming back to the apparently rapid change in the turbulent crossover for ${\rm Ek} < 8 \times 10^{-6}$, this feature is unexpected and entirely unexplained.

The last feature of note in our phase diagram in fig.\ \ref{fig:PhaseDiagram} is the range of possible existence for a scaling range with Nu $\sim$ Ra$^3$Ek$^4$.  The assumption of the marginal stability arguments leading to that scaling is that the stability of thin thermal boundary layers of width $\delta$ is the same as for the bulk rotating convection problem, namely that one can write \cite{KingJFM2012}
\begin{equation}
{\rm Ra}_c^\delta = \frac{\rm Ra}{2} \left(\frac{\delta}{d}\right)^3 = B {\rm Ek}_\delta^{-4/3} = B {\rm Ek}^{-4/3} \left(\frac{\delta}{d}\right)^{8/3}
\end{equation}
where $B = 8.4$ and the sub/superscript $\delta$ denotes the evaluation using 
$\delta$ rather than $d$. From this relationship, one solves for $\delta$ to yield $\delta /d = 2^3 B^3 {\rm Ra}^{-3}{\rm Ek}^{-4}$, which implies Nu$=d/(2\delta) \sim {\rm Ra}^3{\rm Ek}^4$. For this to hold, however, one needs Ra$_c^\delta > C$ where $C \gtrsim 1000$, i.e., that the boundary layer feels the effect of rotation rather than being effectively a non-rotating stability problem. In other words, Ra$_c^\delta \sim$ Ek$_\delta^{-4/3}$ is only valid for Ek $<< 1$ which is violated as $\delta$ decreases because Ek$_\delta \sim \delta^{-2}$.  The resulting expression of the absolute upper limit of self consistency of this argument results from solving for Ra/Ra$_c$ above to obtain
\begin{equation}
{\rm Ra/Ra}_c \lesssim  {\rm Ek}^{-1/6}.
\end{equation}
This condition is line F in fig.\ \ref{fig:PhaseDiagram}.  Based on this estimate, it seems unlikely that one could observe this regime for Ek$ > 10^{-5}$, with a solid decade of scaling only possible for Ek$ < 10^{-9}$ and Ek$ < 10^{-15}$ for two decades!

The phase diagram in fig.\ \ref{fig:PhaseDiagram}  suggests the following: 1) The transition from buoyancy dominated turbulent convection to rotation influenced turbulent convection depends sensitively on Pr and approximately scales as Ro. 2) The transition from rotation-influenced to rotation-dominated convection is best described by a transition relationship ${\rm Ra}_t = 0.5 {\rm Ek}^{-7/4}$ with little or no Pr dependence. 3) The available range of Ra/Ra$_c$ is insufficient to observe geostrophic turbulence scaling \cite{JulienPRL12} or Ra$^3$Ek$^4$ scaling \cite{KingJFM2012} for Ek $> 10^{-5}$ which includes almost all of the data taken for water with Pr $\approx 5$ 
\cite{Rossby69,Zhong93,LE-PRL97,KSNHA09,ZSCVLA-PRL09,LE-PRE09}.  4) Although our results for Pr = 0.7 lie on approximately the same transition line to geostrophic turbulence (line C) as the results at higher Pr $> 5$, there may be another branch of transitions (line E in fig.\ \ref{fig:PhaseDiagram}).  5) To observe a full decade of geostrophic turbulence scaling will require Ek $< 10^{-7}$ whereas a similar range for a possible ${\rm Ra}^3{\rm Ek}^4$ range would need Ek $< 10^{-9}$. There are thus many experimental and numerical challenges that need to be addressed to further characterize and extend the fascinating problem of rotating thermal convection.

%Acknowledgements: 
We acknowledge important conversations with Keith Julien, Antonio Rubio and Geoffrey Vasil concerning predictions of numerical simulations.  Contributions by Ecke were funded by the National Nuclear Security Administration of the U.S. Department of Energy at Los Alamos National Laboratory under Contract
No. DE-AC52-06NA25396. We acknowledge the Elettra Synchrotron Laboratory, Trieste, for providing technical and logistical assistance and, in particular, we wish to thank Mr. Piergiorgio Tosolini.

%\bibliographystyle{apsrev4}
%\bibliography{RotConv2013}

\end{document}